\documentclass[conference, 10pt]{IEEEtran}
\usepackage{cite}

\usepackage[pdftex]{graphicx}
\graphicspath{/figs/}
\DeclareGraphicsExtensions{.pdf,.jpeg,.png, .eps, .svg}
\usepackage{amsmath, amssymb}

\usepackage[table]{xcolor}
\usepackage{bm}

\usepackage{float}
\usepackage[caption = false]{subfig}
\usepackage{algorithm}
\usepackage{algorithmicx}
\usepackage{algpseudocode}
\usepackage{pifont}
\usepackage{xcolor}

\usepackage{array}
\usepackage{url}
\usepackage{parskip}
\usepackage{layout}

\usepackage{mathtools}

\algdef{SE}[DOWHILE]{Do}{doWhile}{\algorithmicdo}[1]{\algorithmicwhile\ #1}%

\DeclareMathOperator*{\argmax}{arg\,max}
\newcommand{\Real}{\operatorname{\mathbb{R}}}

\newcommand{\norm}[1]{\left\lVert#1\right\rVert}

\newcommand{\chan}{\operatorname{\mathbf{H}}}
\newcommand{\pathgain}{\operatorname{\alpha_{\ell}}}

\newcommand{\azimuth}[1]{\operatorname{\theta_{\ell}^{#1}}}
\newcommand{\elevation}[1]{\operatorname{\phi_{\ell}^{#1}}}
\newcommand{\vander}[2]{\operatorname{\mathbf{a}_{N_{#2}}(#1)}}
\newcommand{\vect}[1]{\operatorname{\mathbf{#1}}}
\newcommand{\complex}{\operatorname{\mathbb{C}}}
\newcommand{\steering}[1]{\operatorname{\mathbf{A}(#1)}}
\newcommand{\conjsteering}[1]{\operatorname{\mathbf{A}^*(#1)}}
\newcommand{\chanstack}{\operatorname{\tilde{\mathbf{H}}}}
\newcommand{\w}{\operatorname{\mathbf{w}}}

\newcommand{\f}{\operatorname{\mathbf{f}}}
\newcommand{\Lmax}{\operatorname{L_{\text{{max}}}}}
\newcommand{\ue}{\operatorname{u}}

\begin{document}

\makeatletter
\newcommand{\newlineauthors}{%
  \end{@IEEEauthorhalign}\hfill\mbox{}\par
  \mbox{}\hfill\begin{@IEEEauthorhalign}
}
\makeatother

\title{Massive {MIMO} Beam Management in Sub-6 GHz {5G NR}}
\author{\IEEEauthorblockN{Ryan M. Dreifuerst,~\IEEEmembership{Student Member,~IEEE, }}%
	\IEEEauthorblockA{\textit{Dept. of Electrical and Computer Engineering}\\
		\textit{The University of Texas at Austin}\\
		ryandry1st@utexas.edu
	}
    \and
	\IEEEauthorblockN{Robert W. Heath Jr.~\IEEEmembership{Fellow,~IEEE, }}%
	\IEEEauthorblockA{\textit{Dept. of Electrical and Computer Engineering}\\
		\textit{North Carolina State University}\\
		rwheathjr@ncsu.edu
	}
	\and
	\IEEEauthorblockN{Ali Yazdan,~\IEEEmembership{Member,~IEEE, }}%
	\IEEEauthorblockA{\textit{Facebook Inc.}\\
		ayp@fb.com
	}

	\thanks{Ryan M. Dreifuerst is with The University of Texas, Austin, TX 78712 USA (ryandry1st@utexas.edu). 
	        Robert W. Heath Jr. is with North Carolina State University, Raleigh, NC 27695 (rwheathjr@ncsu.edu).
	        Ali Yazdan is with Facebook, Menlo Park, CA 94025 (ayp@fb.com).
	       This work was supported in part by Meta Platforms Inc.}% <-this % stops a space
}

\maketitle
\bstctlcite{IEEEexample:BSTcontrol}
	
\begin{abstract}
    Beam codebooks are a new feature of massive multiple-input multiple-output (M-MIMO) in 5G new radio (NR). Codebooks comprised of beamforming vectors are used to transmit reference signals and obtain limited channel state information (CSI) from receivers via the codeword index. This enables large arrays that cannot otherwise obtain sufficient CSI. The performance, however, is limited by the codebook design. In this paper, we show that machine learning can be used to train site-specific codebooks for initial access. We design a neural network based on an autoencoder architecture that uses a beamspace observation in combination with RF environment characteristics to improve the synchronization signal (SS) burst codebook. We test our algorithm using a flexible dataset of channels generated from QuaDRiGa. The results show that our model outperforms the industry standard (DFT beams) and approaches the optimal performance (perfect CSI and singular value decomposition (SVD)-based beamforming), using only a few bits of feedback. 

    % Massive multiple-input multiple-output (M-MIMO) is an important technology for future mobile networks. Prior work has largely considered multi-user M-MIMO for millimeter-wave bands while ignoring sub-6GHz bands. As a result, it is unclear how effective M-MIMO and multi-user technologies can be for mobile broadband networks. Multiple data streams can be transmitted on spatially multiplexed systems to efficiently serve multiple users by balancing the directivity and spatial dimensionality of the precoder. Accurately producing the correct beams, however, requires perfect channel state information (CSI) or relies on codebook-based beam training. We show that site-specific codebooks can be learned to adapt to user distributions and RF environments. We pose the problem in the most uninformative situation, initial access, where cells have no initial knowledge of users. Our proposed network uses beamspace observations to adopt site-specific synchronization signal (SS) burst patterns that enable extremely low feedback beam alignment. Our results show that we can achieve comparable performance to the perfect-CSI case while requiring only a few bits of feedback.
    
\end{abstract}

\section{Introduction}
    Support for MIMO in 5G has been rethought with an eye towards providing unifying support at both lower and higher frequencies. Going beyond 4G, a beam management protocol has been introduced to enable the use of larger arrays without the need for explicitly estimating the channel. Traditionally, channel state information is obtained by estimating the channel from reference signals for every logical element, with a maximum of $32$ ports in 5G release 16 \cite{3gppTS38.211}. The beam management protocol in 5G allows for massive, fully digital architectures that have the flexibility of large antenna arrays without excessive feedback \cite{AndrewsBMin6G2021}. To maximize the potential for M-MIMO, codebooks should be designed that capture both user distribution and the environment characteristics. Such information is difficult to capture analytically, but we have shown that machine learning is a powerful tool for optimize wireless networks in such settings \cite{DreifuerstEtalCCO}. Due to the large dimensionality and complex underlying relationships, data-driven machine learning has potential to help design and optimize M-MIMO codebooks.
    
	\begin{figure}[ht!]
	    \centering
	    \includegraphics[width=3.45in]{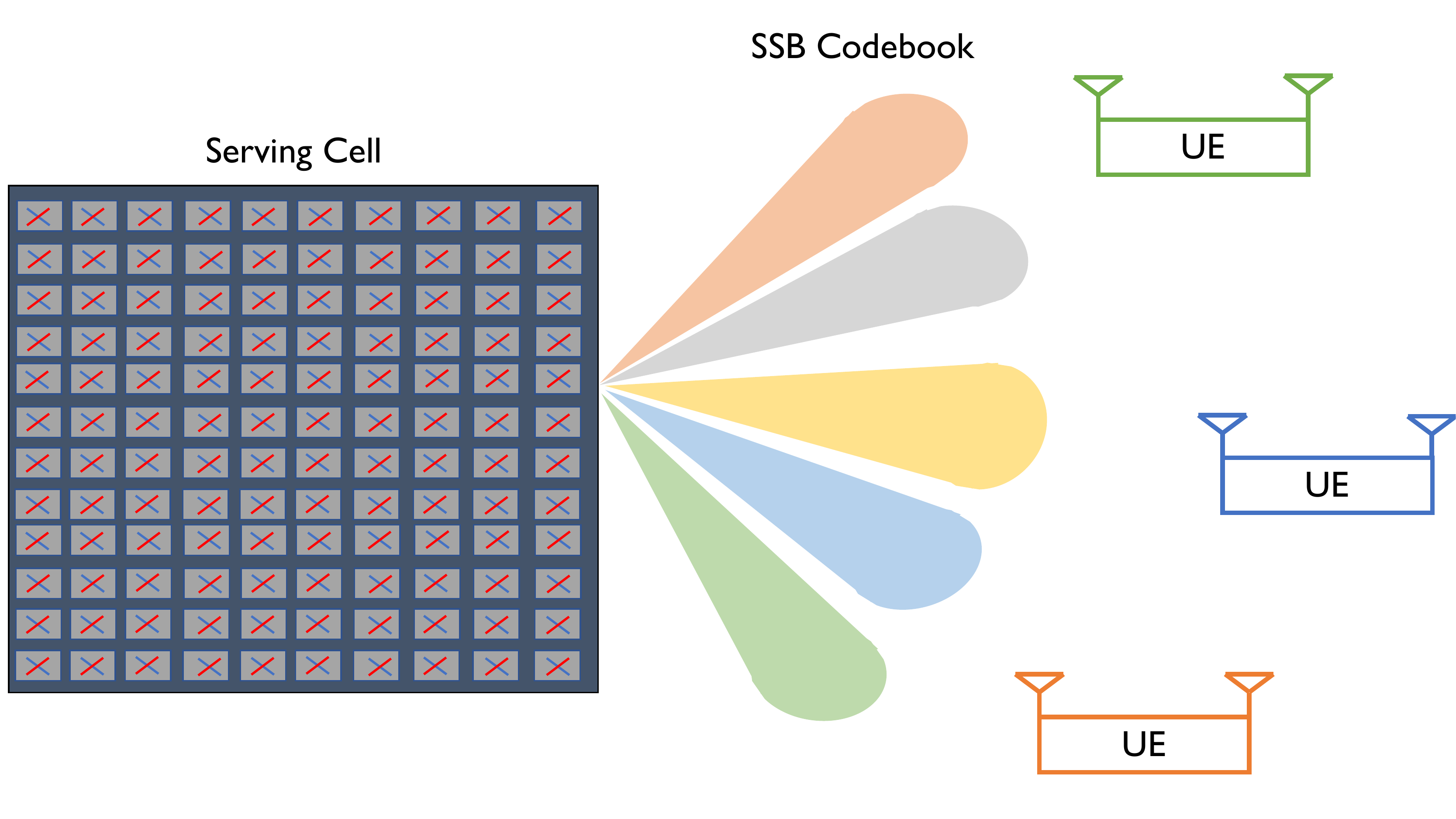}
	    \caption{The SSB codebook is used to transmit beamformed reference signals for limited feedback and synchronization with user equipment (UEs).}
	    \label{fig: system_image}
	\end{figure}

	There is much work on beam training and limited feedback; we review here some select references that are most related to our proposed work on machine learning-based beam-training for M-MIMO in 5G. In one line of work, hierarchical codebook design was investigated \cite{XiaoCodebook2016} for analog and hybrid arrays. Beam switching based on gradient descent methods has been shown to reduce the complexity and feedback in millimeter-wave systems \cite{LiBeamTraining2013}. In a different direction, AI/ML can be incorporated to aid the beam training process \cite{WangCScodebook2021, MyersEtAlDeepLearningBasedBeamAlignment2020, XiaDLFrameworkMISOBF2020, ShafinEtalRLBroadcastBeams2020}.
% 	The design of codebooks can also be integrated with receiver location \cite{Y.WangEtAlMmWaveVehicularBeamSelection2019} to recommend subsets of beams to try. 
	Recently, \cite{WangCScodebook2021} considered the design of compressive codebooks based on the learned channel statistics for the purpose of beam training. In \cite{MyersEtAlDeepLearningBasedBeamAlignment2020}, the concept was extended by replacing the key network parameters with a neural network structure inspired by autoencoders. The focus, however, was primarily on millimeter-wave and did not consider M-MIMO or lower frequency operation. A framework was proposed in \cite{XiaDLFrameworkMISOBF2020} for deep learning in MISO beamforming using a convolutional network. Others have looked at deep reinforcement learning approaches to capture UE distributions for sectored base stations \cite{ShafinEtalRLBroadcastBeams2020}. It was shown that full dimension codebook beamforming can be learned to match time-varying UE distributions \cite{ShafinEtalRLBroadcastBeams2020}. The network was focused on maximizing the number of connected users, however, rather than high data rate connections. In general, these investigations have only considered single stream or single antenna UEs, and often allow for significant feedback \cite{WangCScodebook2021, MyersEtAlDeepLearningBasedBeamAlignment2020, XiaDLFrameworkMISOBF2020}. We see the lack of realistic multi-user MIMO investigations with limited feedback as an important research direction due to the complexity and dimensionality involved. Given that machine learning has been shown to be an effective method for incorporating relationships between user density and mobility with beam management, our investigation into sub-6GHz M-MIMO beamforming is well-situated to address this gap.
	
    In this paper, we propose a neural network (SSB-Encoder) architecture, which improves the initial access beams based on learned user distribution and wireless environment characteristics. Our algorithm is trained in a supervised manner to direct beams toward active users, while still broadly covering regions where users may become active. We use the QuaDRiGa \cite{JaeckelQuadriga} framework to generate realistic wireless channels and process the data into an initial access scenario. We show that our algorithm learns to project beams that trade off between directivity and coverage, while also producing beams that cover distinct regions. In site-specific testing, our algorithm approaches the optimal performance of a perfect CSI system while only needing a few bits of feedback per user.

\section{System model}
    We begin our investigation with an overview of the analytical model and synchronization signal block (SSB) beamforming used in initial access. Afterward, we define the problem formulation and metrics of interest for our framework. Throughout this paper, we will limit the problem to a single cell with multiple UEs each equipped with multiple antennas and assume the UE and base station do not obtain channel estimates.

    \subsection{Channel model}
    We model the system so that it is representative of real-world conditions to analyze a realistic beam training scenario for sub-6GHz M-MIMO. For this reason, we use spatially consistent channels with fully digital architectures. We do not explicitly specify TDD or FDD because our work does not rely on channel reciprocity. That being said, our algorithm is based on very limited feedback, so the starkest benefits will be for an FDD system. We assume a half-wavelength spaced uniform planar array (UPA) with $N_X \times N_Y$ elements at the base station in a downlink broadcast transmission. The steering vectors $\steering{\theta, \phi} \in \complex^{N_x \times N_y}$ are the Kronecker product of the steering vectors of a uniform linear array (ULA) in each dimension $\vect{a}(\theta, \phi) = \vect{a}(\theta) \otimes \vect{a}(\phi)$, and the responses are given by the Vandermonde vector and Kronecker product
    \begin{align}
        \vander{\theta}{} =&\ [1, e^{j \pi \cos{\theta}}, e^{j 2 \pi \cos{\theta}}, ...,\ e^{j(N-1) \pi \cos{\theta}}]^T  \label{eqn: vander} \\
        \steering{\theta, \phi} =&\ \vander{\theta}{x} \otimes \vander{\phi}{y} \in \complex^{N_x\times N_y}. \label{eqn: steering}
    \end{align}
    We further assume that the channel is constant over a symbol period and narrowband. We leave evaluation of wideband channels to future work. The channel is defined for the angular pairs between receiver $\ue$ and the transmitter, $(\azimuth{R}, \elevation{R}), (\azimuth{T}, \elevation{T})$ and complex gain ${\pathgain}$ for a set of $L_p$ paths as
    \begin{align}
        \chan^{(\ue)} &=\ \sum_{\ell=1}^{L_p} \pathgain \steering{\azimuth{R}, \elevation{R}} \otimes \conjsteering{\azimuth{T}, \elevation{T}} \label{eqn: MIMO} \\
         &\in \ \complex^{(N^R_X \times N^R_Y) \times (N^T_X \times N^T_Y)}.
    \end{align}
    The same result can be achieved by viewing the channel as the tensor product of the two dimensional azimuth channel response and elevation channel response. We further restrict the system to only consider a ULA at the receiver, equivalent to setting $N^R_Y=1$, and simplifying the channel components to a three dimensional tensor. The UE's array is assumed to be properly oriented with respect to the base station, such that a receiver could achieve the full array gain with an appropriate combining strategy. Such an assumption is partially feasible through fully digital, multi-panel arrays, although we anticipate UE orientation will be valuable for future investigations. We now remove the superscripts from the receiver and transmitter values, assuming $N_R$ is the full set of antennas at the receiver. In many future steps, it will be useful to stack the channel as a 2D response of the full $N_T = (N_X N_Y)$ such that
    \begin{equation}
        \chanstack_{i, j}^{(\ue)} = \chan_{i, n, m}^{(\ue)} \quad \forall i \in \{0, 1, ... N_R\}, j=n N_X + m.
    \end{equation}
    This representation allows for analyzing planar systems with traditional MIMO techniques.

	\subsection{Initial access}
		Initial access is the first process a UE must go through when connecting to a mobile network. In 5G NR, a cell may initiate the initial access period at regular intervals of $\{5, 10, 20, 40, 80, 160\}$ms \cite{Giordani2019} to control the frequency that a UE must be active and providing feedback to the network. During this period, the cell will transmit SSBs that contain primary and secondary synchronization signals (PSS, SSS), as well as demodulation reference signals (DMRS) \cite{Giordani2019}. These SSBs are precoded using a specific codeword. Depending on the cell carrier frequency and subcarrier spacing, a cell may transmit up to $L = \{1, 4, 8, 64\}$ beams in a burst, and all cells must transmit at least one beam. During transmission, the UE will measure the Reference Signal Received Power (RSRP) and report the index of the beam with the highest RSRP. There are two key aspects of the RSRP as a reporting metric. 1) RSRP does not account for interference, and 2) if the UE is equipped with multiple antennas, it may either receive the signal via all antennas with one or more receiving weight combining strategy, or limit the receiving to the first antenna. 
    
        We can now define the two metrics of interest: the received signal reference power (RSRP) and the cosine similarity. The RSRP is one of the primary metrics that the receiver will measure during initial access and is used for determining the channel quality index (CQI) and SSB index (SSBRI). The base station will use this value to determine the strength of the signal, the code rate to be used, and the overall beamformer. The cosine similarity is a metric for the base station to evaluate how similar two beamformers are. Ideally, each SSB has dissimilar beamformers to maximize the independent coverage area and directivity, however, the distribution of the UEs can lead to conflicting goals between maximizing the RSRP and ensuring beams are dissimilar.
        % In single-receive antenna settings, maximizing the RSRP effectively maximizes the signal strength and the achievable capacity with the UE. The capacity is not directly dependent on the RSRP in a MIMO or multi-user setting, however, because the RSRP is only relative to the strongest stream and so there is no multi-stream feedback beyond the rank indication without additional CSI reporting.
        
        The RSRP is one of the basic channel quality metrics and is determined by measuring the received power during a given reference signal. In the case of initial access, the reference signal is given by the demodulation reference signal (DMRS), which is a known training sequence for both the transmitter and receiver. The measurement of the RSRP is performed after applying a receive combining filter, $\w$, for the $i^{\text{th}}$ SSB beamformed signal $\f_i$. We will aggregate the transmit/receive power, gain factors, and large scale channel SNR into a single value, $\gamma_{\ue}$. The issue of UE beamforming (selecting $\w$) is not a focus in this investigation and is left to future work. Instead, we will assume the UEs perform MRC perfectly with respect to the SSBs and channel, thereby coherently combining across all of the $N_R$ antennas. In fully digital arrays the UE would use the DMRS signals for estimation and filtering to achieve a similar setting. The resulting RSRP is with additive noise $\vect{n}$ is obtained as
        \begin{align}
            p_i^{(\ue)} &=\ \frac{\gamma_{\ue}}{N_T}  \norm{\chanstack^{(\ue)} \f_i + \vect{n}}^2. \label{eqn: rsrp}
        \end{align}
        
        % \begin{align}
        %     p_i^{(\ue)} &=\ \frac{\gamma_{\ue}}{N_T}  \norm{\chanstack^{(\ue)} \f_i}^2. 
        % \end{align}
        Note that \eqref{eqn: rsrp} is for a specific UE $(\ue)$ and SSB $(i)$, so the total feedback, $\vect{p}, \vect{m}$, is a vector of the maximum RSRP values for each of the UEs and the associated SSB indices
        \begin{align}
            \vect{p} &=\ \{\max_i{p_i^{(\ue)}}\}_{\ue=0}^{U-1} \label{eqn: Prsrp} \\
            \vect{m} &=\ \{\argmax_i{p_i^{(\ue)}}\}_{\ue=0}^{U-1}. \label{eqn: Irsrp}
        \end{align}
        The other metric, cosine similarity, is a classic method for comparing two vectors. We use the cosine similarity to compare how much two beams overlap, which impacts how effective the downlink precoding will be. In the ideal setting, each of the top $\Lmax$ UEs would select a different beam and each beam would cause no interference with other beams. Due to multipath propagation that is difficult for the base station to determine without additional feedback, so we instead evaluate how the directive beamforming pairs overlap in similarity. The cosine similarity between beams $\vect{f}_a$ and $\vect{f}_b$ is evaluated as
        \begin{align}
            \Delta_{\text{sim}} = \frac{|\vect{f_a}^*\vect{f_b}|}{\norm{\vect{f_a}} \norm{\vect{f_b}}}. \label{eqn: cos_sim}
        \end{align}
        The cosine similarity is measured between each combination of beams to determine beam overlap of the entire codebook.
        
        It is possible for the UE to estimate and feedback the channel as well. Unfortunately, the number of CSI ports is limited to $32$ \cite{3gppTS38.211}, so some sort of decimation or upscaling is necessary for M-MIMO arrays to obtain CSI for every port. This also incurs an overhead penalty that is orders of magnitude larger than the limited SSB feedback. Instead, we restrict the setting to the minimum feedback according to \eqref{eqn: Prsrp} and \eqref{eqn: Irsrp}. In such a setting, the base station is tasked with A) proposing the next set of SSBs to use, and B) determining the best precoder to maximize the capacity of the active UEs. In this paper, we will focus on the first task; an evaluation of the precoder performance would require an extensive investigation under different environment conditions and scheduler algorithms.

\section{Proposed SSB-Encoder network}
    We now define the problem of learning the codebook to use in initial access for an M-MIMO 5G NR cellular network. The goal is to use the limited feedback from the UEs to predict the next SSB codebook that can serve the users, while ensuring new UEs can also be served. Ultimately, the learning algorithm should propose a set of coefficients $\vect{F} \in \complex^{\Lmax \times N_X \times N_Y}$ based on the feedback $\vect{p}, \vect{m}$ that maximizes the RSRP of the UEs. 
    % Ideally, the proposed vectors should also be sufficiently orthogonal to limit the interference between UEs during a subsequent downlink transmission.
    Ideally, the proposed vectors should also be sufficiently orthogonal such that a suboptimal successive interference cancellation \cite{CoverSIC1972} policy can achieve reasonable capacity.
    
	\begin{figure}[!t]
	    \centering
	    \includegraphics[width=3.45in]{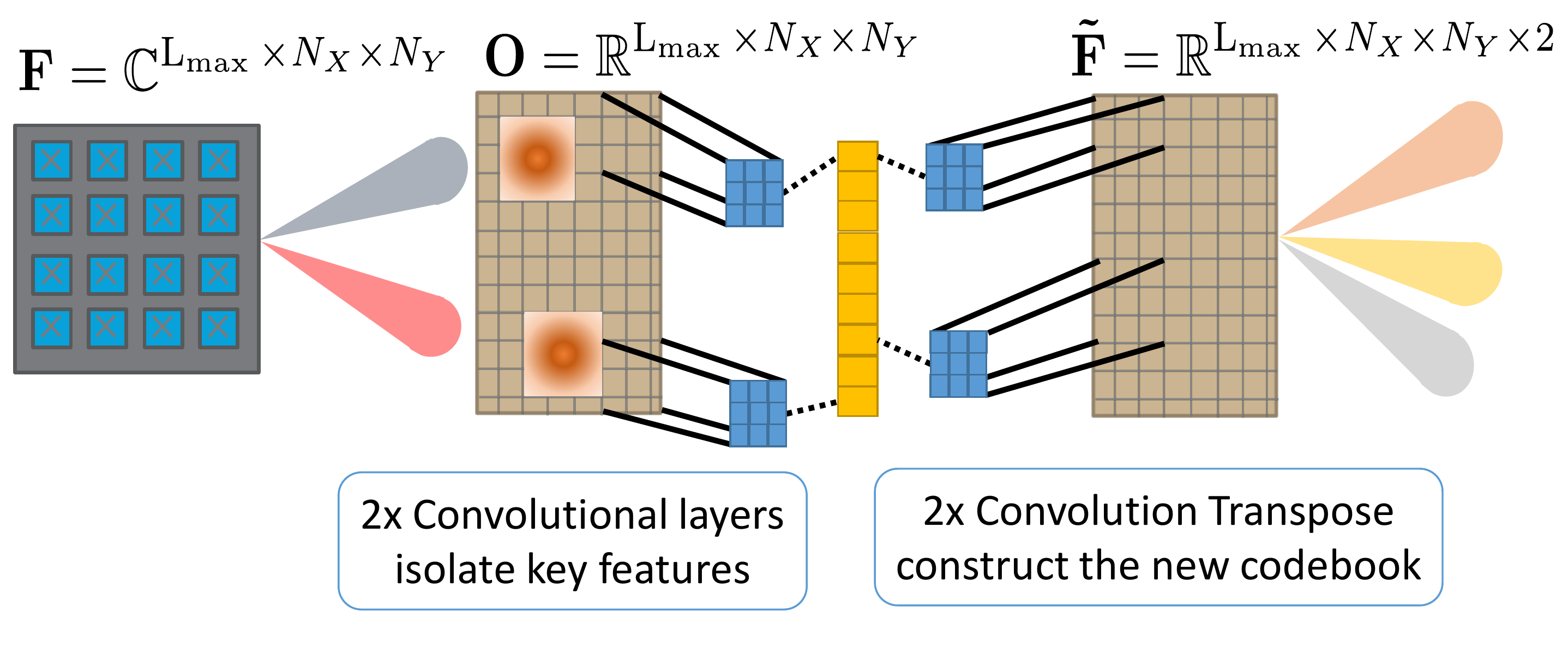}
	    \caption{Visualization of the beamspace observation and SSB encoder architecture. The beamspace converts the beams to virtual beam directions and the autoencoder architecture isolates the features and constructs a new codebook.}
	    \label{fig: model}
	\end{figure}
	
    We use a virtual beamspace observation as the input, which translates the codewords into a universal, directional grid. The base station prepares the observation by first discretizing the angular grid space into $N_X$ azimuth points and $N_Y$ elevation points for each SSB. Then it determines which beams were reported back from the active UEs and which points in the angular grid space are in the top $A_{\text{max}}$dB of the associated beams. These regions are set to $1$, smoothed with a Gaussian kernel of size $4\times4$, multiplied by the reported RSRP, and normalized to have a Frobenius norm of $1$ for each beam. If multiple UEs report the same beam, then the corresponding grids are summed up. The result is an observation matrix, $\vect{O} \in \Real^{\Lmax \times N_X \times N_Y}$ that provides a natural translation from the vector feedback to a beamspace representation.

    Intuitively, if there was no consistency between samples, i.e. if the radio frequency environment changed in both large- and small-scale manners and the UE were uniformly spaced over the beamspace, then there would be little that can be learned or improved upon from the basic DFT codebooks that dominate traditional systems. In a realistic setting, however, both UE distributions (azimuth and elevation trajectories) and spatial consistency lead to information that can be used to improve the results. We propose an autoencoder architecture with the observation matrix, shown in Figure \ref{fig: model} to learn this information. The encoder structure uses two convolutional layers at the encoder side with zero-padding and ReLU activation functions, with the inverse (transpose convolution) at the decoder. The final decoder layer has only a linear activation and produces outputs of size $\vect{\tilde{F}} \in \mathbb{R}^{\Lmax \times N_X \times N_Y \times 2}$, which corresponds to the real and imaginary components of the resulting SSB set. To start the SSB-Encoder, we create an observation model where each DFT beam was reported once with equal RSRP. This produces the most uninformed setting to iteratively progress from, although the algorithm has no trajectory or historical information to rely on.
    
    We build a dataset to train the model by evaluating a random subset of the channels, calculating the SVD of the channel for each user, and selecting the best $L$ vectors based on the largest singular values. The selected beamforming vectors are the ideal SVD-based output that we train the model to produce, given only the observation matrix obtained using wide DFT beams. We set $A_{\text{max}}=6$dB to include more information than just the half-power beamwidth, but avoid regions that may arise due to sidelobes of the array pattern. 
    % We expect a multi-step timeseries approach to improve the results further, but leave such an exploration for future work.
    The model is trained using an Adam optimizer with learning rate reduction and early stopping to minimize the mean squared error between the SVD SSB beamformers and the proposed $\vect{\tilde{F}}$. Data is split into $180000$ samples for training and $20000$ samples for validation. Upon testing, we generate $2000$ new channel sets.

\section{Simulation results} \label{sec: sim_results}
    First, we define the simulation setup and channel generation. We use the QuaDRiGa \cite{JaeckelQuadriga} channel simulator for the initial channel realizations and post process the data to fit the initial access situation. After defining the simulation setup, we report the RSRP results showing how the proposed SSB-Encoder performs. We also look at the distribution of beam choices reported by the UEs and the codebook similarity. 
    % we report the site-specific results, which are when the same distribution of data is used for training and testing. We then look at the site agnostic results, by evaluating the algorithm with a new distribution of data. These two scenarios show how the site-specific learning is learning information implicit within the RF environment.
    
    \subsection{Simulation setup}
        We simulate random realizations from a single sector base station and UEs distributed according to one of two possibilities with probability $\{0.3, 0.7\}$. In the first case, UEs are classified as stationary and scattered uniformly over the region. Alternatively, UEs are placed along a specified roadway with normally distributed speeds of $25$m/s and a standard deviation of $5$m/s. Each UE may be line of sight (LOS), non-LOS (NLOS), or a combination as it moves according to the wireless model. The base station is equipped with $N_X=N_Y=8$ antennas, and the UE has $N_R=4$ antennas. We use a carrier frequency $f_c = 3.5$GHz with the 3GPP 3D UMi model \cite{Mondal3GPP3D2015} and 3D radiation patterns. This choice of carrier frequency leads to up to $\Lmax=8$ SSBs per burst. The UEs are allowed to move for two seconds while being sampled every $5$ms. The full channels for all of the antenna pairs for each UE at every timestep are saved to be processed into the initial access format.
        % The scattering objects and wireless parameter distributions are held constant throughout the simulations.
        
        The post processing first randomly selects a starting time sample, $T_{\text{start}}$ and a random number of active UEs, $U_{\text{active}} \sim \mathbb{U}(4, 12)$, are selected from the channels at the starting time. At each timestep, a UE will drop into or drop out of the network with probability $0.2$. This represents the chance that new users become active or that the scheduler assigns new users to join the network. UEs that remain active have correlated channel patterns, while new UEs can appear at any location based on the current timestep and UE classification. This ensures the network remains able to adequately cover the entire region and is not entirely focused on previously-active UEs.

    \subsection{RSRP results}
        Because the data is random and episodic, we look at the RSRP as a function of the simulation index or as a function of the beam index. The simulation indices correspond to $5$ms intervals, and after $20$ intervals an entirely new set of UEs and timeslots are chosen. In Figure \ref{fig: RSRP}, we show the resulting RSRP of our algorithm compared to purely wide-DFT beams and a system with perfect CSI at the transmitter using an SVD approach (CSIT-SVD). The DFT beams are generated so that the range of potential beams is split into $4$ azimuth beams and $2$ elevation beams for a total of $\Lmax=8$ beams. Effectively, this splits the coverage into $4$ primary-coverage regions and $4$ cell-edge regions. We can see that our algorithm bridges the gap between the two extremes: optimal, perfect CSI beamforming and uninformed wide DFT beams. On average, our algorithm recovers more than half of the performance difference between the DFT and CSI-SVD approaches with only a few bits of feedback. 
        
		\begin{figure}[!t]
		    \centering
		    \includegraphics[width=3.45in]{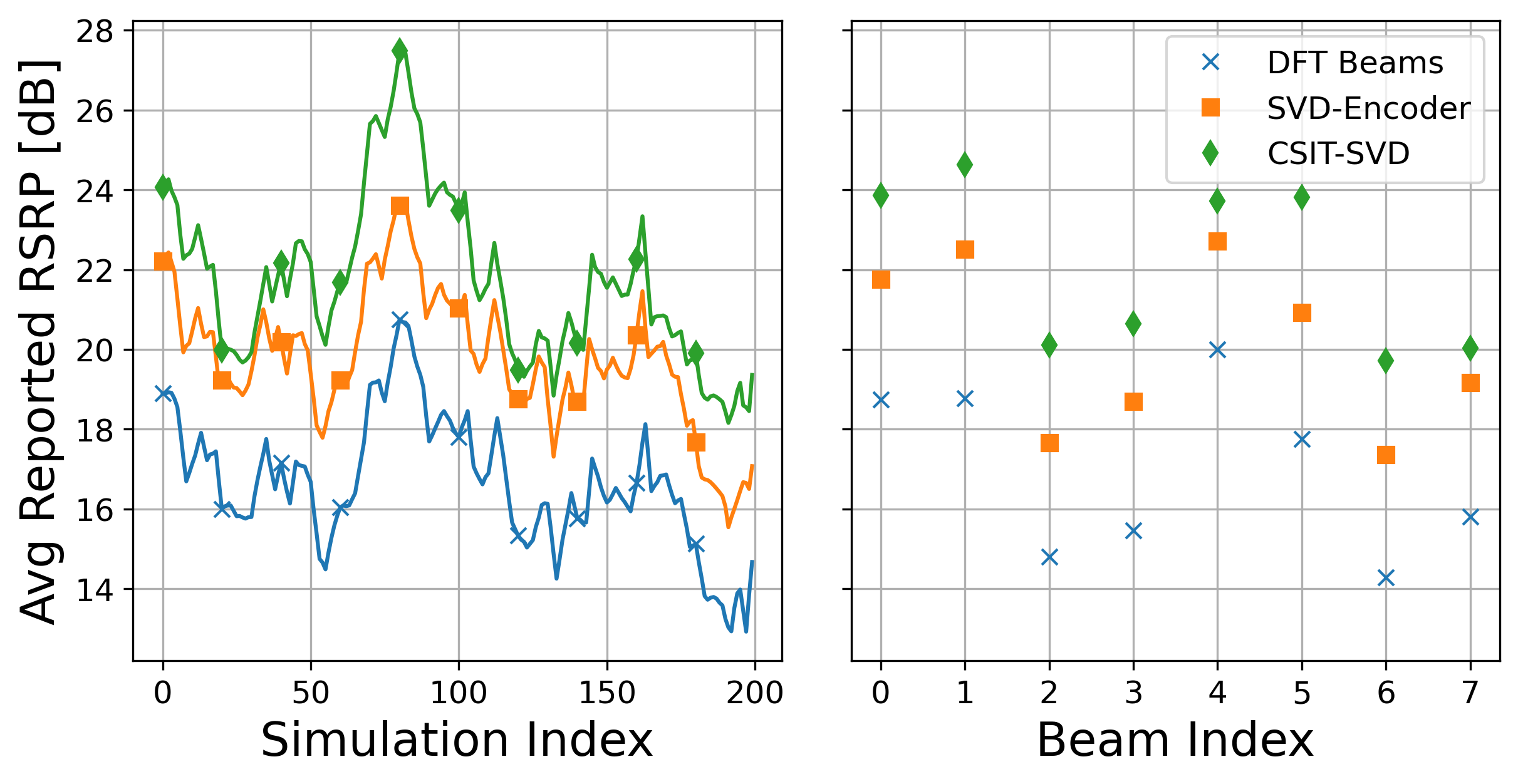}
		    \caption{The RSRP results of the SSB codebook with averaging over all active UEs in the first plot and over time in the second. We smooth the data according to a $20$ point averaging filter in the first plot to improve the readability. }
		    \label{fig: RSRP}
		\end{figure}

    \subsection{Beam selection}
        The distribution of the reported beams, $\vect{m}$, is important because the ability of the system to spatially separate the UEs is directly affected by the choice of beams and the beam overlap. The normalized histogram of using DFT beams and the initial set of our algorithms beams are shown in Figure \ref{fig: distributions} after $10,000$ samples. We can see that the DFT beams rarely use the odd numbered beams, which have less downtilt and correspond to cell edge locations. In contrast, our algorithm has a more uniform split over the beam choices, improving the separability of the UEs. While this is helpful, the UE separability also depends on how much overlap occurs between the beams. In the case of wide DFT beams, as used here, there will naturally be some overlap for primary/cell-edge beams. We show a heatmap of the cosine similarity in Fig. \ref{fig: similarity} between the two sets of proposed beams. 
        % We can see that although there is more overlap in our algorithm's codebook, only one pair of beams, beams $2$ and $7$, have high similarity.
        
		\begin{figure}[!t]
		    \centering
		    \includegraphics[width=3.1in]{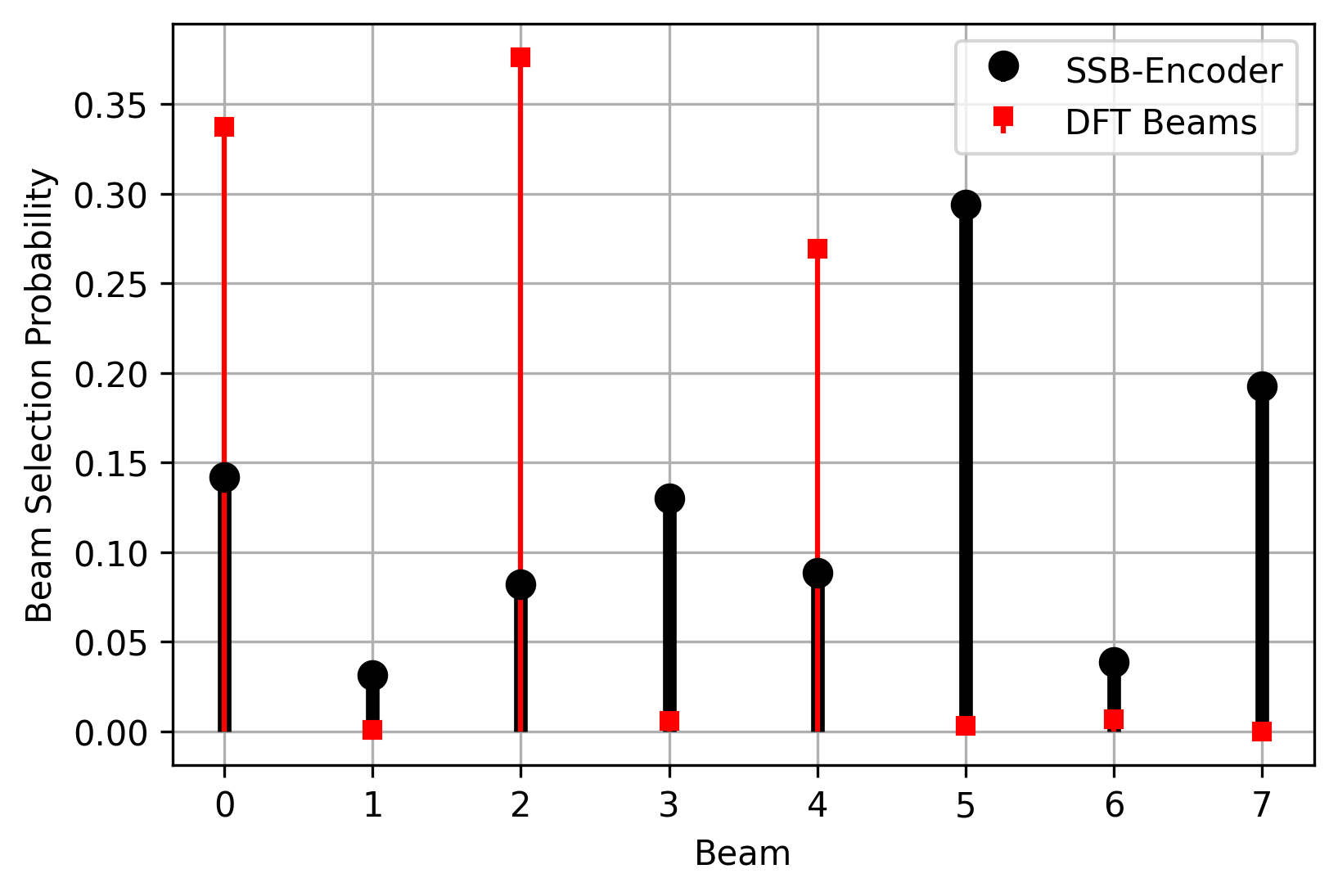}
		    \caption{A comparison of the beam selection distributions from $10,000$ samples. It can be seen that the DFT beams rarely use the odd valued beams, while our encoder distribution is closer to uniformly distributed. This distribution aids the system in separating the UEs in the beamspace domain.}
		    \label{fig: distributions}
		\end{figure}

		\begin{figure}[!t]
		    \centering
		    \includegraphics[width=3.2in]{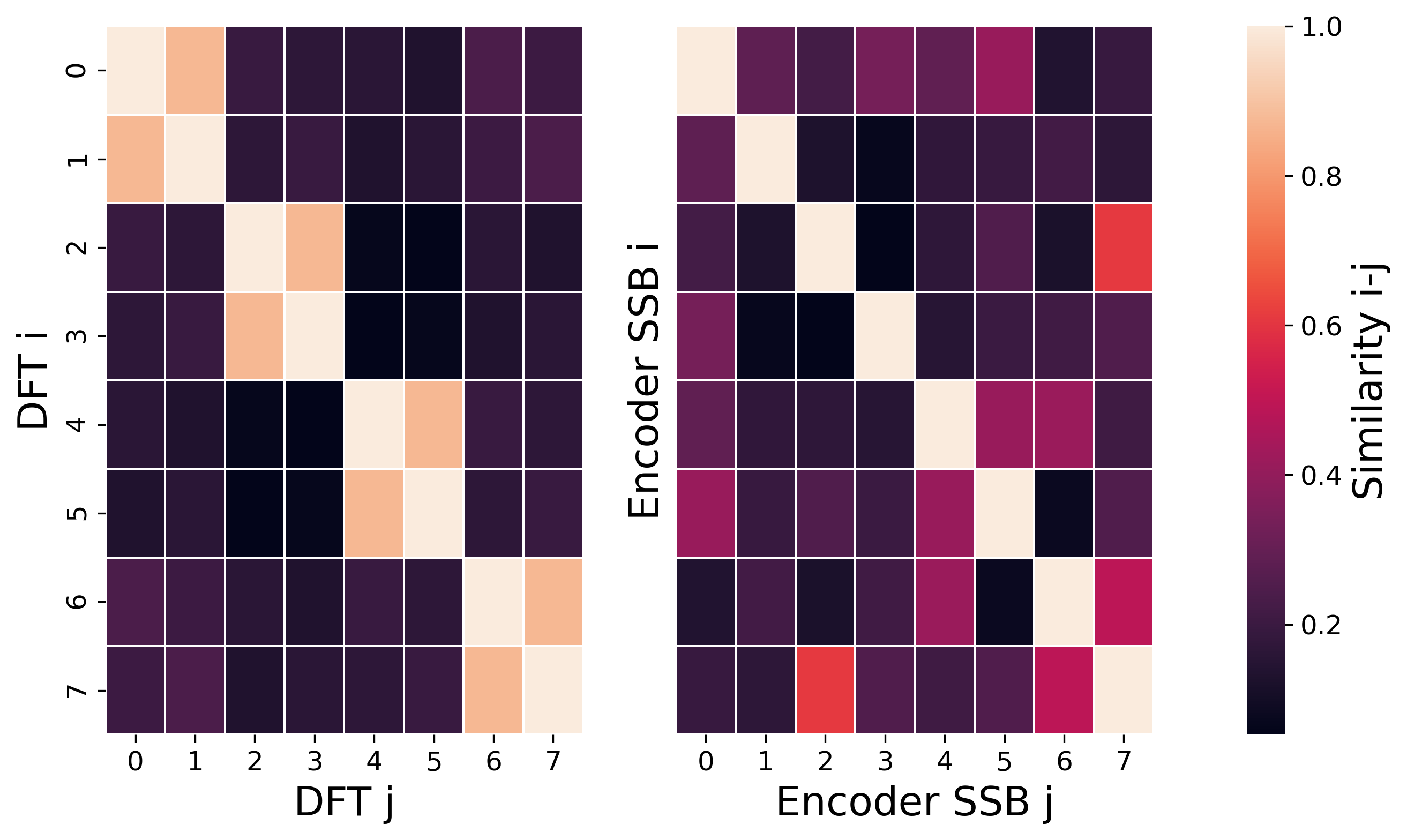}
		    \caption{A heatmap of the cosine similarity for our algorithm's initial codebook and wide DFT codebook. We can see that the similarity of our codebook is slightly larger than DFT beams, but generally never reaches as high similarity as is seen by the azimuth-aligned DFT beams, i.e. the even-odd pairs.} 
		    \label{fig: similarity}
		\end{figure}
    
		Finally, we plot two beam projections from the SSB-Encoder in Figure \ref{fig: beams}. It can be seen that the algorithm appears to learn non-overlapping beams while attempting to cover the whole projection space. We can also see that, unlike millimeter-wave beams, the beams are not exceptionally directional. In fact, the beams only reach about $12$dBi, whereas a directive beam could reach almost $23.5$dBi. 
		
		\begin{figure}[!t]
    	    \centering
    	    \subfloat[ ]{\includegraphics[width=3.15in]{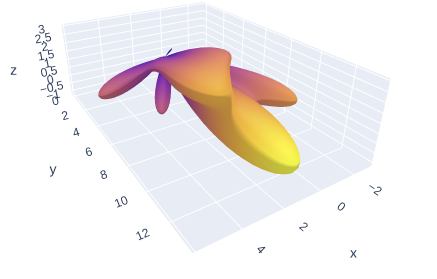}}
    	    
    	    \subfloat[ ]{\includegraphics[width=3.1in]{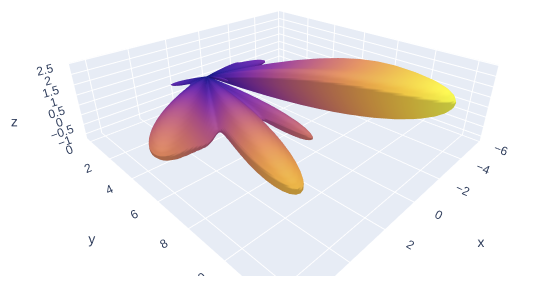}}
    	    \caption{Two SSB beamforming examples from the SSB-Encoder. The two beams appear to cover distinct regions, as is expected based on the heatmap in Figure \ref{fig: similarity}. The maximum beamforming gain is about $10$-$12$dB for each SSB.}
    	    \label{fig: beams}
    	\end{figure}

\section{Conclusion}
    In this paper, we have presented a novel framework for learning initial access beams for sub-6GHz 5G NR. Using limited feedback and beamspace observations, our algorithm is able to bridge the performance gap between perfect CSI systems and generic DFT codebook beamforming. The algorithm uses an autoencoder type architecture to learn the RSRP-maximizing SVD-based beams in a narrowband channel model. Using the dynamic codebook generated by the SSB-Encoder, the SSB performance is improved by more than $3$dB with only a few bits of feedback in the current 5G framework. In future work, we will expand the investigation to include wideband and millimeter-wave channels.

\bibliographystyle{IEEEtran}
\bibliography{IEEEabrv, references, heath_refs_all}
\end{document}